\documentclass[conference]{IEEEtran}
\IEEEoverridecommandlockouts
\pdfoutput=1
\usepackage{cite}
\usepackage{amsmath,amssymb,amsfonts}
\usepackage{algorithmic}
\usepackage{graphicx}
\usepackage{textcomp}
\usepackage{xcolor}
\def\BibTeX{{\rm B\kern-.05em{\sc i\kern-.025em b}\kern-.08em
    T\kern-.1667em\lower.7ex\hbox{E}\kern-.125emX}}
\usepackage{tikz}
\usepackage{textcomp}
\usepackage{hyperref}
\usepackage{lipsum}

\newcommand\copyrighttext{
  \footnotesize \textcopyright~2021 IEEE. Personal use of this material is permitted. Permission from IEEE must be obtained for all other uses, in any current or future media, including reprinting/republishing this material for advertising or promotional purposes, creating new collective works, for resale or redistribution to servers or lists, or reuse of any copyrighted component of this work in other works. DOI: \href{https://ieeexplore.ieee.org/document/9681801}{10.1109/EnT50460.2021.9681801}}
\newcommand\copyrightnotice{
\begin{tikzpicture}[remember picture,overlay]
\node[anchor=south,yshift=10pt] at (current page.south) {\fbox{\parbox{\dimexpr\textwidth-\fboxsep-\fboxrule\relax}{\copyrighttext}}};
\end{tikzpicture}
}    
    
\begin{document}

\title{A Study of the Impact of the Contention Window on the Performance of IEEE 802.11bd Networks with Channel Bonding\\
	\thanks{The research has been carried out at IITP RAS and supported by the Russian Science Foundation (Grant No 20-19-00788, https://rscf.ru/en/project/20-19-00788/)}
}

\author{\IEEEauthorblockN{Viktor Torgunakov$^{*\dagger}$, Vyacheslav Loginov$^*$, Evgeny Khorov$^{*\dagger}$}
	$^*$Institute for Information Transmission Problems of the Russian Academy of Sciences, Moscow, Russia \\
	$^\dagger$Moscow Institute of Physics and Technology, Dolgoprudny, Russia \\
	e-mail: $\{$torgunakov, loginov, khorov$\}$@wireless.iitp.ru}

\maketitle

\begin{abstract}
	Nowadays, Vehicle-To-Everything (V2X) networks are actively developing. Most of the already deployed V2X networks are based on the IEEE~802.11p standard. However, these networks can provide only basic V2X applications and will unlikely fulfill stringent requirements of modern V2X applications. Thus, the IEEE has launched a new IEEE~802.11bd standard. A significant novelty of this standard is channel bonding. IEEE~802.11bd describes two channel bonding techniques, which differ from the legacy one used in modern Wi-Fi networks. Our study performs a comparative analysis of the various channel bonding techniques and a single-channel access method from IEEE~802.11p via simulation. We compare them under different contention window sizes and demonstrate that the legacy technique provides the best quality of service in terms of frame transmission delays and packet loss ratio. Moreover, we have found a quasi-optimal contention window size for the legacy technique.   
\end{abstract}


\section{Introduction}
\label{sec:intro}
\copyrightnotice
Vehicle-To-Everything (V2X) networks can highly decrease the number of fatalities and enhance driving comfort. It can be achieved, for example, by using applications that warn a driver about an unexpected appearance of a pedestrian or a sudden road accident. As V2X networks are becoming more efficient, in the future, they are expected to ensure the work of fully autonomous vehicles~\cite{V2X-offloading}.  

A key radio access technology that provides V2X networks is Dedicated Short Range Communications (DSRC). This technology relies on a set of standards for Wireless Access in Vehicular Environment (WAVE), where IEEE~802.11p specifies physical and medium access control layers. DSRC networks were designed to operate in the 5.9~GHz frequency band dedicated for V2X networks in Europe, the USA, etc.

Various studies show that IEEE~802.11p networks can fulfill the requirements of basic safety applications if network traffic is moderate~\cite{11p-provide-basic-apps-1, 11p-provide-basic-apps-2}. Such applications utilize periodic message broadcasting with a 1--10~Hz generation rate, require a 50--500~ms end-to-end latency, and a Packet Loss Ratio (PLR) below 10$\%$~\cite{ASTM-basic-apps-PLR, ETSI_applications}. 

However, advanced V2X applications impose stricter requirements, and IEEE~802.11p networks will unlikely satisfy them. For instance, a short-range application for all-around view requires the end-to-end latency and the PLR below 3~ms and 0.001$\%$, respectively, which is the most stringent requirement among all advanced V2X applications~\cite{3gpp_requirements}. Therefore, IEEE has established an IEEE~802.11bd Task Group to develop a new standard to enable modern V2X applications. 

The IEEE~802.11bd standard is designed to enhance transmission reliability, throughput, and transmission range compared with IEEE~802.11p. To achieve it, IEEE~802.11bd introduces a set of novelties. A significant novelty of IEEE~802.11bd is channel bonding, which allows transmission of data in two adjacent channels simultaneously and may noticeably boost the performance of stations (STAs) that use IEEE~802.11bd.

The IEEE~802.11bd draft~\cite{11bd_draft} specifies two channel bonding techniques, which differ from the legacy one used in Wi-Fi since the IEEE~802.11n standard (Wi-Fi~4). Naturally, IEEE~802.11bd does not specify when to use one or another technique. Thus, it raises an issue of selecting the most suitable technique depending on the scenario. 

Although some IEEE~802.11bd features have already been studied in many papers~\cite{germany_11bd_draft, germany_phy_abstraction, ma2021sinr, NGV_review1, NGV_review2, korea_1}, to the best of our knowledge, there is only one paper devoted to the channel bonding techniques in IEEE~802.11bd~\cite{korea_1}. Furthermore, the authors compare the techniques only in terms of the frame transmission delays neglecting the PLR even though the PLR is also a crucial metric. Moreover, the authors consider only the default contention window size. 

This paper addresses these drawbacks and conducts a comparative analysis of the channel bonding techniques through extensive simulations in ns-3 in terms of both the PLR and the frame transmission delay. Moreover, we explore the impact of the contention window size on the performance of the techniques. The influence of the contention window size is investigated because it can significantly enhance the performance of IEEE~802.11p based networks~\cite{adaptive-cw1, adaptive-cw2, adaptive-cw3}. 

The rest of the paper is organized as follows. Section~\ref{sec:methods} describes the channel bonding techniques from IEEE 802.11n and IEEE 802.11bd. In Section~\ref{sec:scenario}, we give a review of a considered scenario and quality of service (QoS) metrics. In Section~\ref{sec:numerical}, we evaluate the performance of the considered techniques via simulation. Section~\ref{sec:conclusion} provides a conclusion.

\section{Channel bonding techniques in IEEE 802.11n and IEEE 802.11bd networks}
\label{sec:methods}

\subsection{IEEE~802.11n channel access methods}
\label{subsec:11n}
The STA that uses IEEE~802.11n can use either Enhanced Distributed Channel Access (EDCA) or a channel bonding.

EDCA is a basic channel access method in modern Wi-Fi networks and implies an operation only in one channel. When using EDCA, a STA performs a backoff procedure before every transmission attempt. Initially, the backoff counter is evenly distributed in the range [0, $CW$], where $CW$ is the current contention window. $CW$ is constant for broadcast traffic. The STA decrements the backoff counter whenever the channel is idle for a time $\sigma$ and suspends the counter when the channel becomes busy. The STA resumes and additionally decrements the counter when the channel is idle for a time $\tau$. $\tau$ equals an Extended Interframe Space (EIFS) if the STA did not successfully decode the transmitted packet when the channel was busy, and an Arbitration Interframe Space (AIFS), otherwise. When the backoff counter reaches zero, the STA transmits the frame.

\begin{figure}[!t]
	\begin{minipage}[h]{1\linewidth}
		\center{\includegraphics[width=1\linewidth]{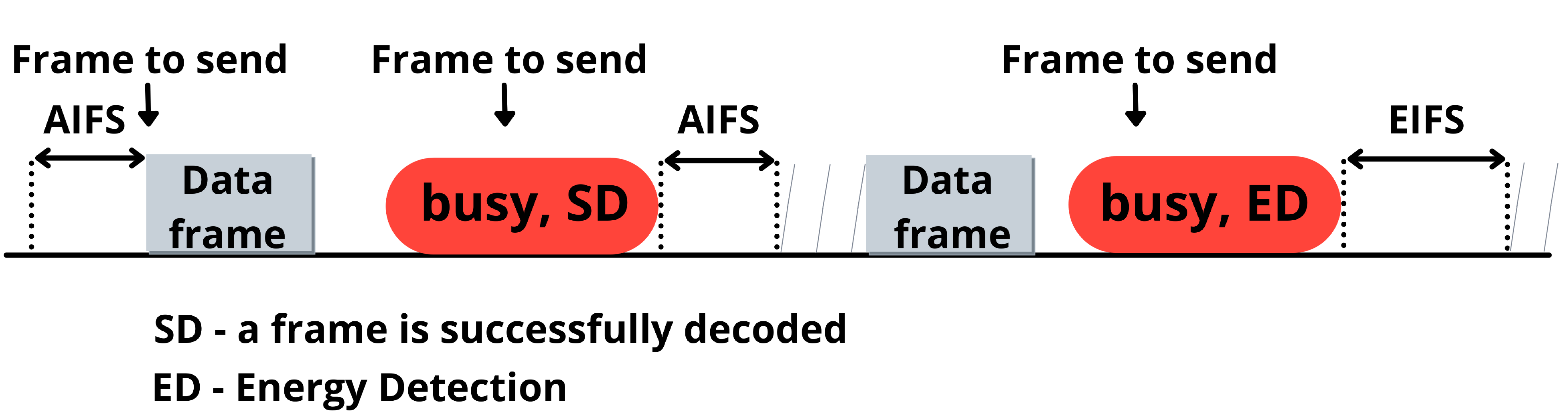} \\ (a)}
	\end{minipage}
	\begin{minipage}[h]{1\linewidth}
		\center{\includegraphics[width=1\linewidth]{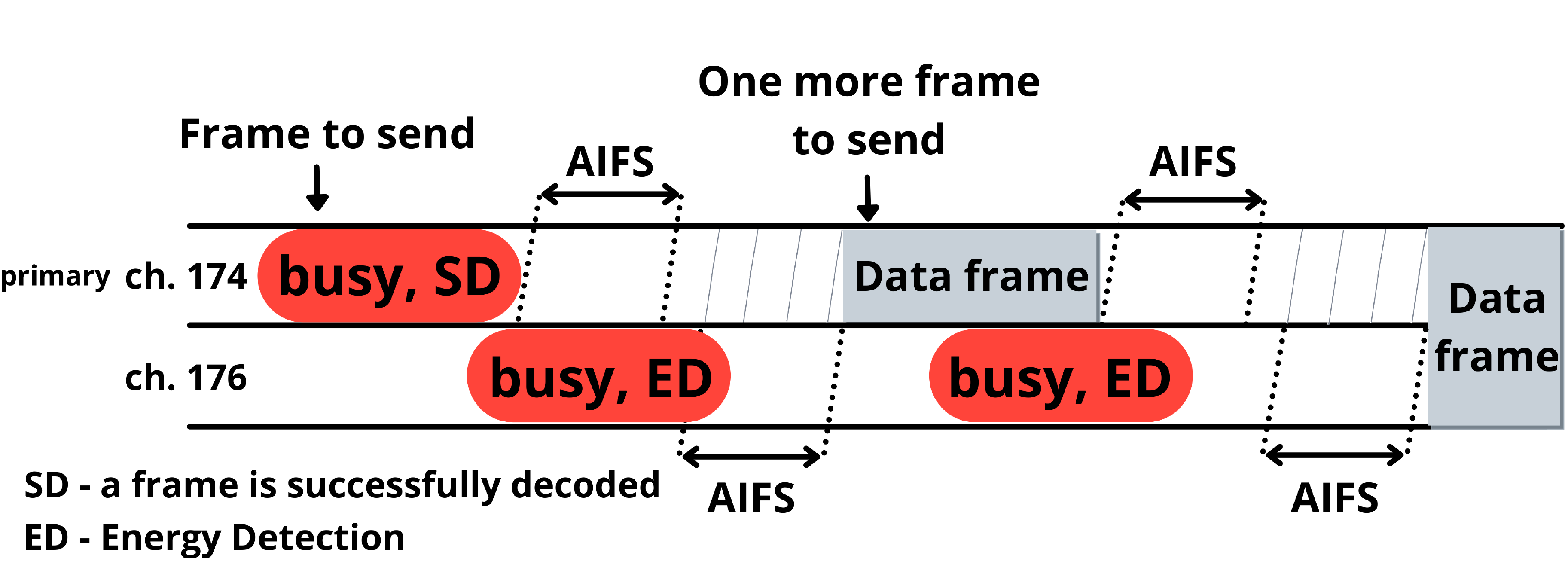} \\ (b)}
	\end{minipage}
	\begin{minipage}[h]{1\linewidth}
		\center{\includegraphics[width=1\linewidth]{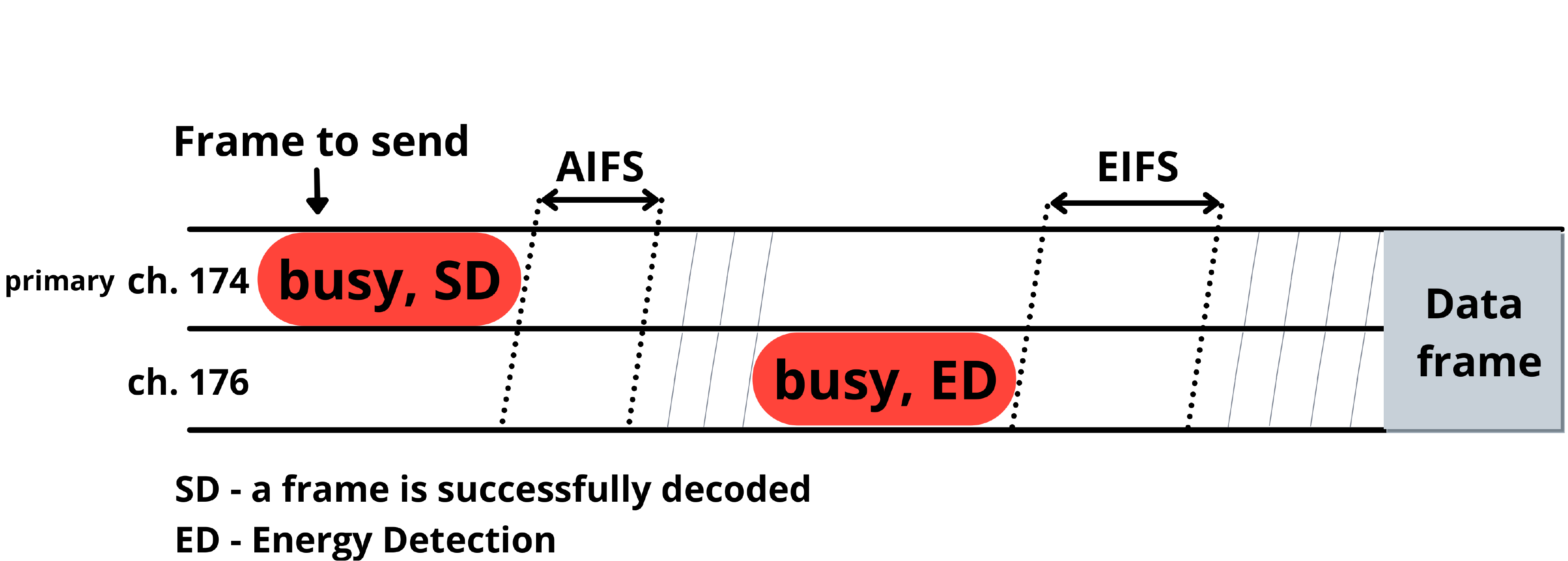} \\ (c)}
	\end{minipage}
	\begin{minipage}[h]{1\linewidth}
		\center{\includegraphics[width=1\linewidth]{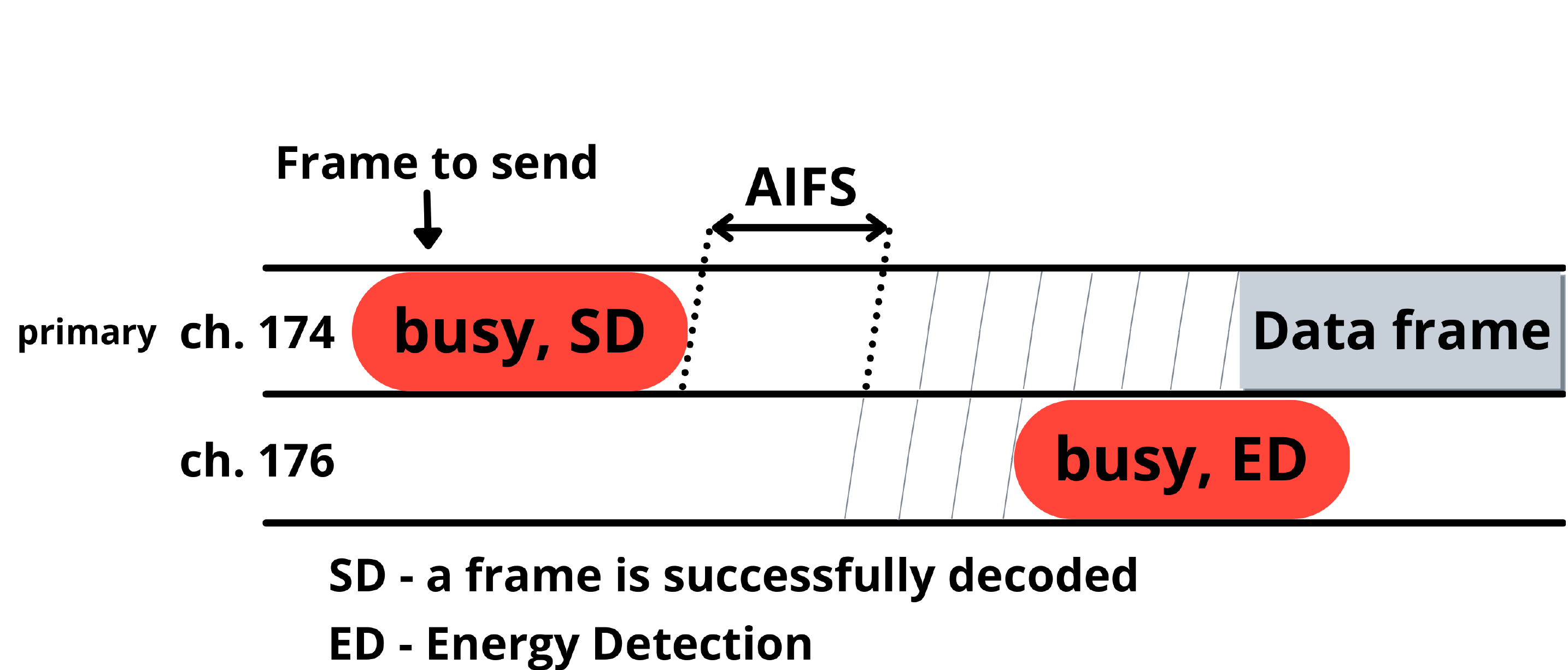} \\ (d)}
	\end{minipage}
	\caption{\label{fig:methods} The operation examples of (a) EDCA; (b)  the technique from IEEE~802.11n; (c) the technique from IEEE~802.11bd without fallback; (d) the technique from IEEE~802.11bd with fallback.}
\end{figure}

A STA that uses IEEE~802.11n splits the wide channel into primary and secondary channels. In the primary channel, the STA counts down the backoff counter and can detect a frame preamble. In the secondary channel, the STA can only determine that the channel is busy if the energy in the channel is above the energy detection threshold. The state of the secondary channel does not have an influence on the backoff procedure in the primary channel.

When the backoff counter reaches zero, the STA checks the state of the secondary channel. If the secondary channel is idle for less than a Point Coordination Function Interframe Space (PIFS), the STA transmits the frame only in the primary channel. Otherwise, it transmits the frame in two channels at once. Figures~\ref{fig:methods}~(a) and~\ref{fig:methods}~(b) provide an example of the channel access methods from IEEE~802.11n operation. 

\subsection{IEEE~802.11bd channel bonding techniques}
\label{subsec:11bd}
A STA that uses IEEE~802.11bd can utilize EDCA, a channel bonding technique with fallback or a technique without fallback. If a STA uses a channel bonding, it splits the wide channel into primary and secondary channels. 

In contrast to IEEE~802.11n channel bonding, a STA that uses the technique from IEEE~802.11bd without fallback suspends the backoff counter if at least one channel becomes busy during the backoff procedure. The STA resumes the backoff counter only when both channels are idle for $T$. $T$ equals EIFS if the STA does not successfully decode a transmission when both channels or the secondary channel are busy last time. Otherwise, $T$ is equal to AIFS. The STA can transmit the frame only in two channels at once.   

There is only one difference between the channel bonding techniques of IEEE~802.11bd with and without fallback. If the secondary channel becomes busy during the backoff procedure when the primary channel is idle, a STA that uses the technique with fallback switches to the primary channel. After that, the STA finishes the backoff procedure in the primary channel and transmits the frame in this channel. After the transmission, the STA returns to the channel access procedure in two channels at once.  
Figures~\ref{fig:methods}~(d) and~~\ref{fig:methods}~(c) illustrate the work of the techniques with and without fallback.

\section{Considered scenario and QoS metrics}
\label{sec:scenario}

\subsection{Available DSRC channels}
\label{subsec:channels}
In~2021, the Federal Communications Commission (FCC) reallocated the 5.9~GHz frequency band, which consisted of seven 10~MHz channels and 5~MHz delimiter and was initially dedicated exclusively for DSRC~\cite{FCC_new}. The FCC authorized unlicensed use in the lower 45~MHz of the band (5.850--5.895~GHz) but retained the upper 30~MHz of the band (5.895--5.925~GHz) for ITS, particularly for the DSRC. Due to this reallocation, some IEEE~802.11p networks deployments have already moved to the upper 30 MHz band~\cite{california_testbed}, and in particular to channels~180 (5.895--5.905~GHz) and 182 (5.905--5.915~GHz), where the channel bonding techniques considered in the paper may be utilized. 

\subsection{Considered Scenario}
\label{subsec:scenario}
In this paper, we consider a scenario where vehicles move along a 1~km long highway~\cite{highway}. The highway has two~sides with four 4~meters wide lanes. The sides are separated from each other by a 25~meters wide median.

There are two types of STAs on the highway: (i) roadside units (RSUs) that use IEEE~802.11p, and (ii) vehicles that use IEEE~802.11bd. 

The number of vehicles on each side of the highway is the same, and the speeds of the vehicles are uniformly distributed from 10 to 30~m/s. The vehicles do not change their line and virtually move at the beginning of the appropriate side upon reaching the end of the line. In turn, the RSUs do not move and are located on both highway sides with a period of 300 meters.

To take into account different network configurations, we consider two cases: symmetric and asymmetric. In the asymmetric case, all the STAs have the same primary channel. In the symmetric case, the STAs on one side of the highway select Channel~182 as primary, whereas all the STAs on another side have primary Channel~180.

In this study, we neglect the effect of adjacent channel interference, as the IEEE~802.11p/bd standards imply stricter requirements on the spectrum mask than other IEEE~802.11 standards.

We load the vehicles with two types of messages. Firstly, the vehicles broadcast Basic Safety Messages (BSMs), which carry information about the size, the acceleration, the speed, and the position of a transmitting vehicle and are widely utilized in the US. The typical generation rate and size of BSMs equal 10~Hz and 250 bytes, respectively~\cite{car_to_car_2020}.

Secondly, the vehicles broadcast Cooperative Perception Messages (CPMs), which are used to share data from the vehicles' sensors. The size of CPMs depends on the number of vehicles in the sensors range $R$. In our study we suppose that the base size of CPMs equals 250 bytes, and each vehicle in range $R = 150$~m  adds 30 additional bytes~\cite{car_to_car_2020, sensors}. Since many of the already deployed safety applications are based on the BSM messages exchange, both 11p and 11bd~STAs should be able to obtain them. Thus, the vehicles transmit BSMs using EDCA. In turn, CPMs are transmitted using the channel bonding techniques because these messages are for emerging applications and are not actively utilized yet.

In turn, the RSUs also broadcast three types of messages. The first two types are Signal Phase and Timing and Map Data (SPaT and MAP) messages used in many real deployments. These messages carry the data about the traffic light signal and the geometry of the intersection near the RSU. STAs transmit SPaT and MAP messages in the same frame, but the SPaT generation rate equals 10~Hz, whereas the MAP generation rate equals 1~Hz. The typical size of the SPaT message alone is equal to 120 bytes, whereas the size of both messages equals 1200 bytes~\cite{california_testbed, car_to_car_2020}. In the symmetric case, RSUs also broadcast WAVE Service Advertisement (WSA) messages, which carry information about the location of transmitting RSU and the channel to switch to. With this information, a vehicle can determine its highway side and tune into the appropriate channel. The generation rate and the size of WSA messages are equal to 1~Hz and 100 bytes, respectively~\cite{1609_3}.

In this paper, we consider the log-distance path loss model. Thus, a path loss is determined as follows: 
\begin{equation}
	PL(l) = PL(l_0) + 10 \gamma \cdot lg \bigg(\frac{l}{l_0} \bigg),
	\label{eq:log_distance}
\end{equation}
where $\gamma =$~2.83 is a path loss exponent, $PL(l_0) =$~44~dB is a reference loss, $l_0 =$~1~m is a reference distance~\cite{brazil_article}. The Modulation and Coding Scheme (MCS) is MCS-1, and the transmission power is equal to 23~dBm~\cite{23_dBm_argumentation}. All STAs in the experiment have the same contention window size $CW$.

\subsection{QoS metrics}
\label{subsec:metrics}
In this study, we used the PLR and a packet transmission delay as basic QoS metrics. The PLR is computed by \eqref{eq:PLR}, where only the STAs on the same side of the highway as the transmitter are taken into account. 
\begin{equation}
	PLR_{m} = 1 - \frac{\sum\limits_{j=1}^{N_{m}} S^{j}_{m}}{\sum\limits_{j=1}^{N_{m}} F^{j}_{m}},  
	\label{eq:PLR}
\end{equation} 
where $m \in \{\text{``}BSM\text{''}, \text{``}CPM\text{''}, \text{``}SPaT\text{''}\}$. In turn, $N_m$ is the total number of transmitted messages of type $m$.
\begin{itemize}
	\item For $m =$ ``$BSM$'' or $m =$ ``$CPM$'', $F^{j}_m$ is the total number of STAs in the $R_s$ range of the STA that transmitted message $j$; $S^{j}_{m}$ is the total number of STAs that successfully decoded message $j$ and are in the $R_s$ range of the transmitter.
	
	\item For $m = $``$SPaT$'', $F^{j}_{m}$ is the total number of vehicles in the $R_s$ range of the RSU that transmitted SPaT message $j$, $S^{j}_{m}$ is the total number of vehicles that successfully decoded SPaT message $j$ and are in the $R_s$ range of the transmitter.
\end{itemize}

In turn, the packet transmission delay includes the queue waiting time, the duration of the backoff procedure, and the packet transmission time.

In this paper, we compared the channel access methods using the number of unsatisfied STAs. A STA is unsatisfied with a message delivery service if the packet transmission delay is higher than $T_d$ or the PLR is higher than $T_{plr}$. For SPaT, MAP, and BSM messages, $T_{d} = 100$~ms whereas for CPM $T_{d} = 10$~ms. $T_{plr}$ is the same for all messages and equals $10\%$~\cite{ETSI_applications, ASTM-basic-apps-PLR, spat_requirement, 3gpp_requirements}. To satisfy the requirements of all types of traffic together, we used the maximum number of unsatisfied users among SPaT and MAP, BSM, and CPM message delivery services.

\section{Numerical results}
\label{sec:numerical}
For our simulation, we extended the WAVE module from ns-3 by implementing the channel bonding techniques, the aforementioned QoS metrics, SPaT and MAP messages, and CPMs. 

\begin{figure}[!t]
	\begin{minipage}[h]{1\linewidth}
		\center{\includegraphics[width=1\linewidth]{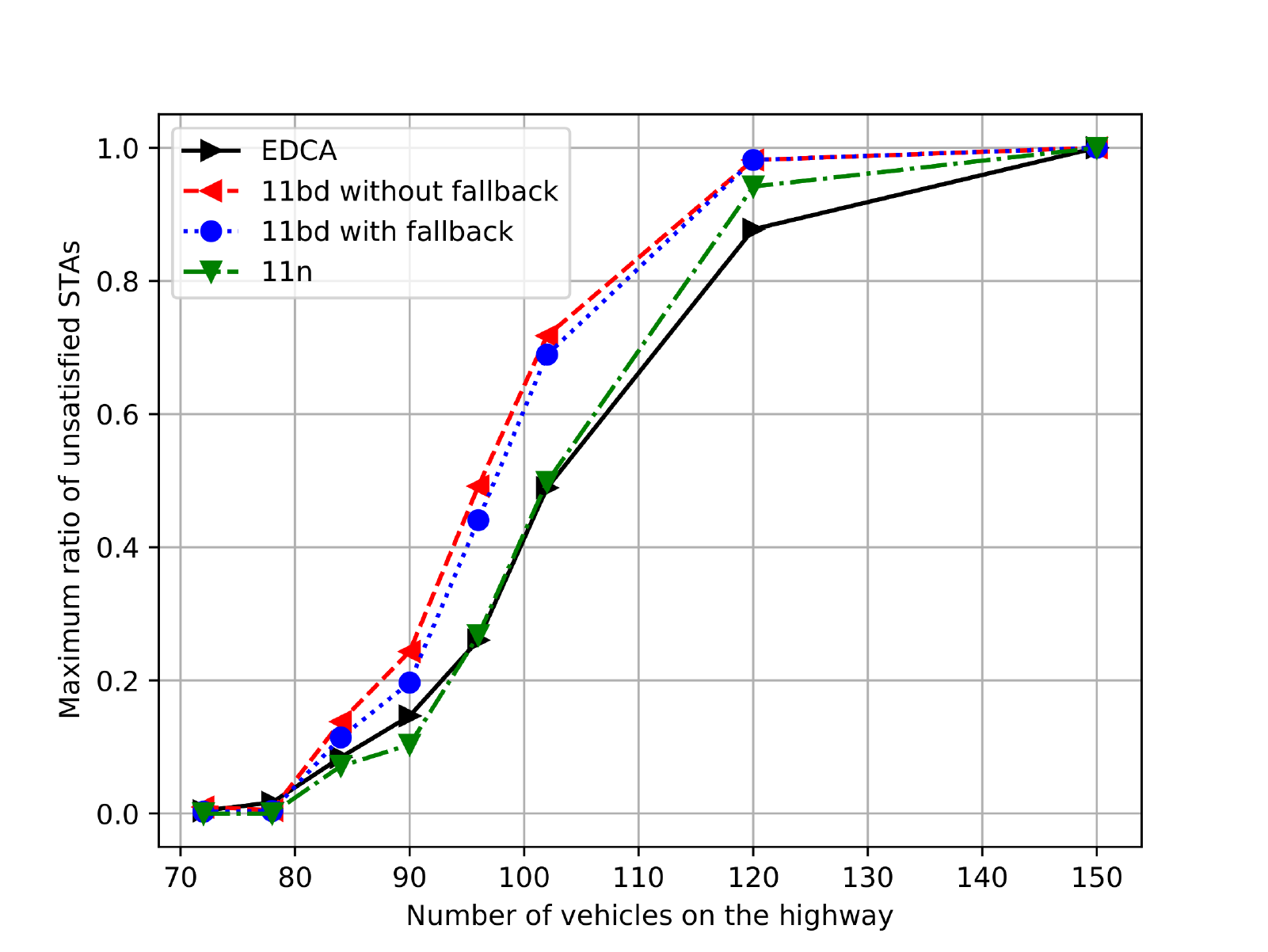}}
	\end{minipage}
	\caption{\label{fig:symm_case_nn} The dependency of the maximum unsatisfied STAs ratio on the number of vehicles on the highway for the symmetric case when $CW = 15$.}
\end{figure}

Fig.~\ref{fig:symm_case_nn} presents the numerical results for the symmetric case and default $CW = 15$.
As we can see, EDCA and the channel bonding technique from IEEE~802.11n outperform the techniques from IEEE~802.11bd in the symmetric case. It is caused by the fact that the PLR of 20~MHz frames is higher than the PLR of 10~MHz frames because a 20~MHz frame can interfere with any other frame, whereas a 10~MHz frame cannot interfere with 10~MHz frames in the adjacent channel.

\begin{figure}[!t]
	\begin{minipage}[h]{1\linewidth}
		\center{\includegraphics[width=1\linewidth]{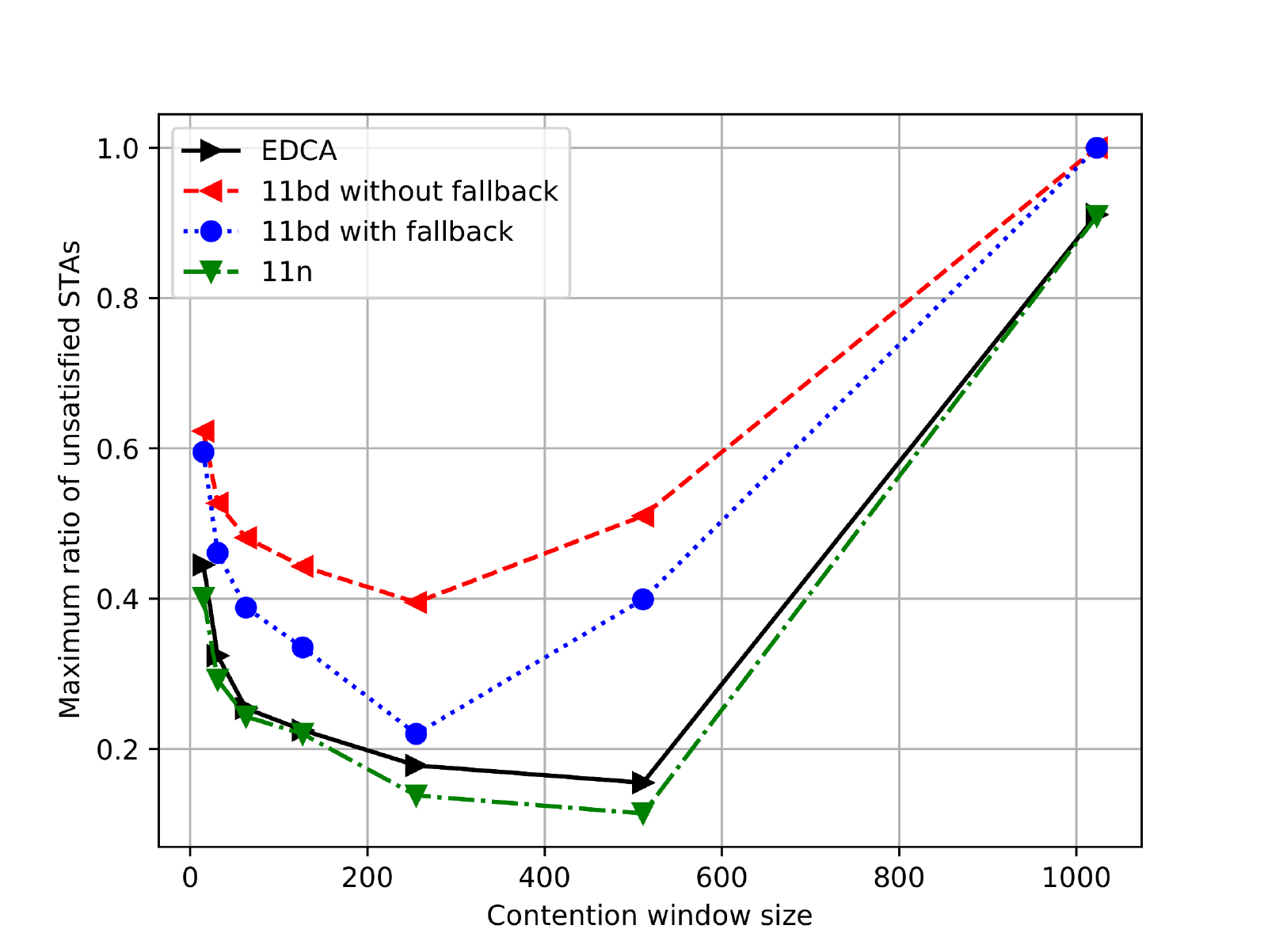}}
	\end{minipage}
	\caption{\label{fig:symm_case_cw} The dependency of the maximum unsatisfied STAs ratio on $CW$ value for the symmetric case when the number of vehicles equals 100.}
\end{figure}

Fig.~\ref{fig:symm_case_cw} provides the results for symmetric case when the number of vehicles on the highway is equal to 100.
The presented results reveal that the optimal $CW$ is smaller for the channel bonding techniques from IEEE~802.11bd and higher for the technique from IEEE~802.11n and EDCA. The optimal $CW$ provides a trade-off between PLR, which is higher when $CW$ is small, and frame transmission delay, which exceeds the threshold for CPM messages when $CW$ is high. The optimal $CW$ is lower for the techniques from IEEE~802.11bd because they perform the channel access procedure in the combined 20~MHz channel, which leads to a noticeable increase in delays.

\begin{figure}[!t]
	\begin{minipage}[h]{1\linewidth}
		\center{\includegraphics[width=1\linewidth]{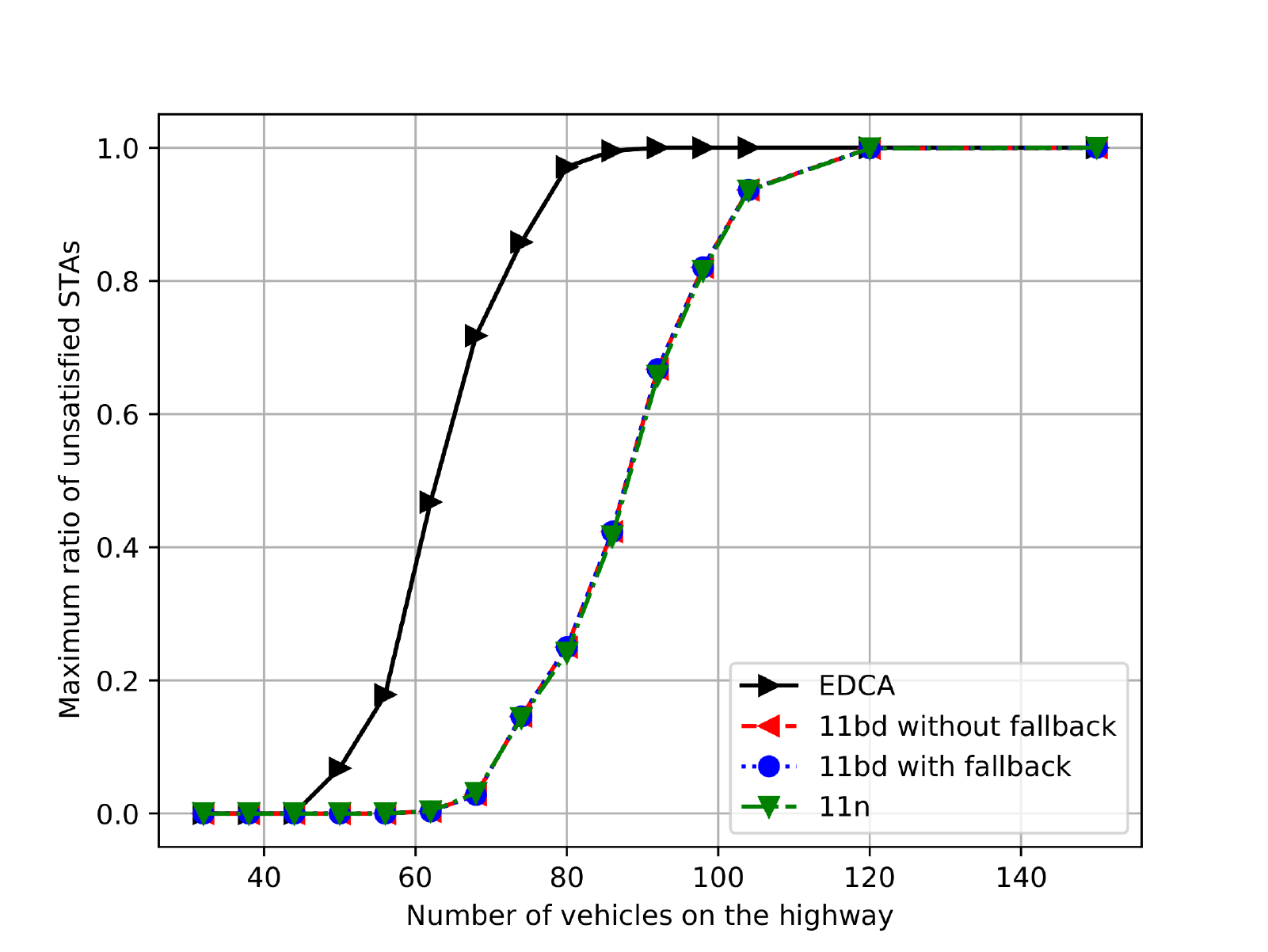}}
	\end{minipage}
	\caption{\label{fig:asymm_case} The dependency of the maximum unsatisfied STAs ratio on the number of vehicles on the highway for the asymmetric case when $CW = 15$.}
\end{figure}
Fig.~\ref{fig:asymm_case} presents the results for the asymmetric case. As we can see, the channel bonding techniques significantly outperform EDCA due to decreased transmission time CPM messages.

\begin{figure}[!t]
	\begin{minipage}[h]{1\linewidth}
		\center{\includegraphics[width=1\linewidth]{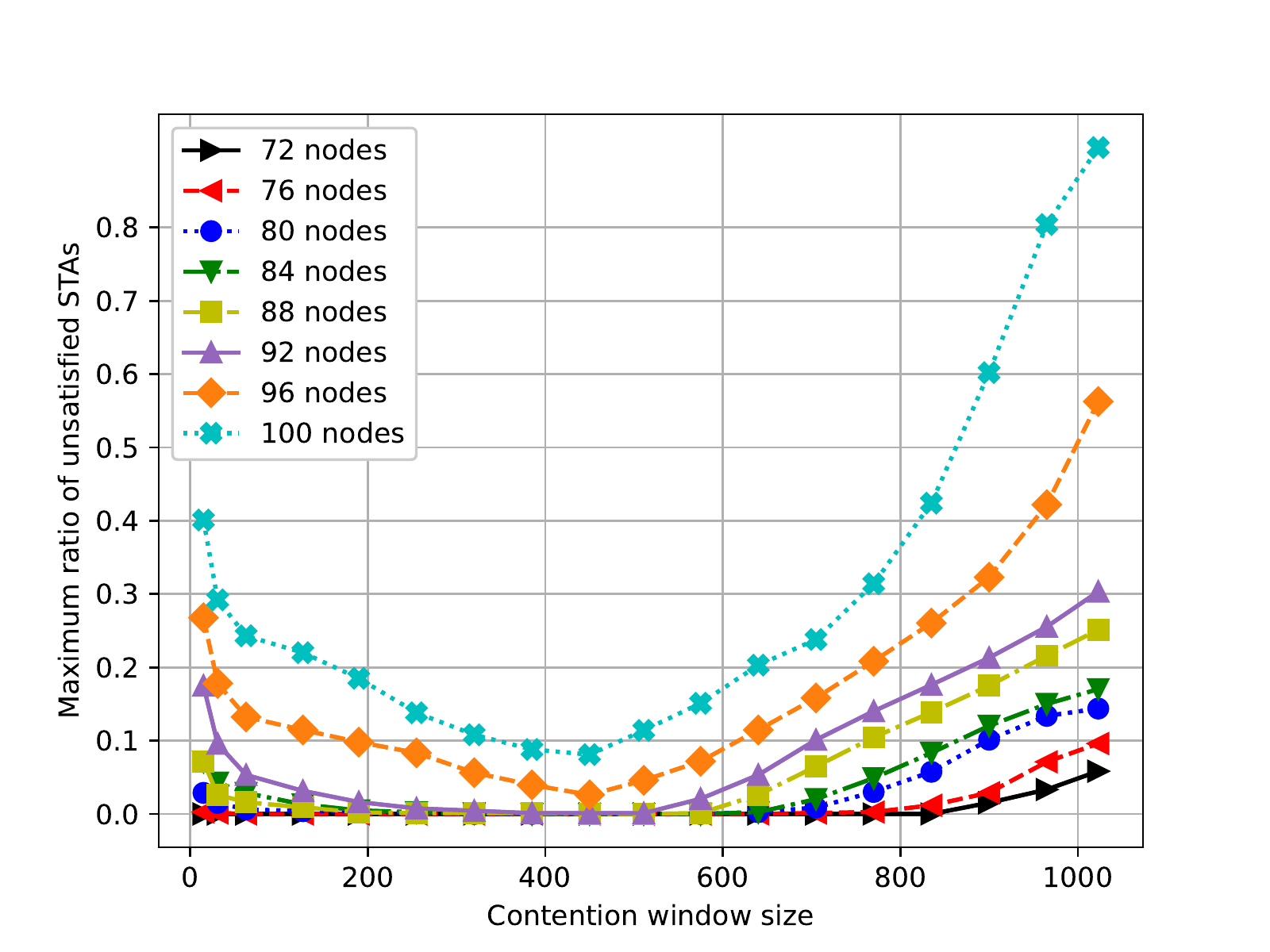} \\ (a)}
	\end{minipage}
	\begin{minipage}[h]{1\linewidth}
		\center{\includegraphics[width=1\linewidth]{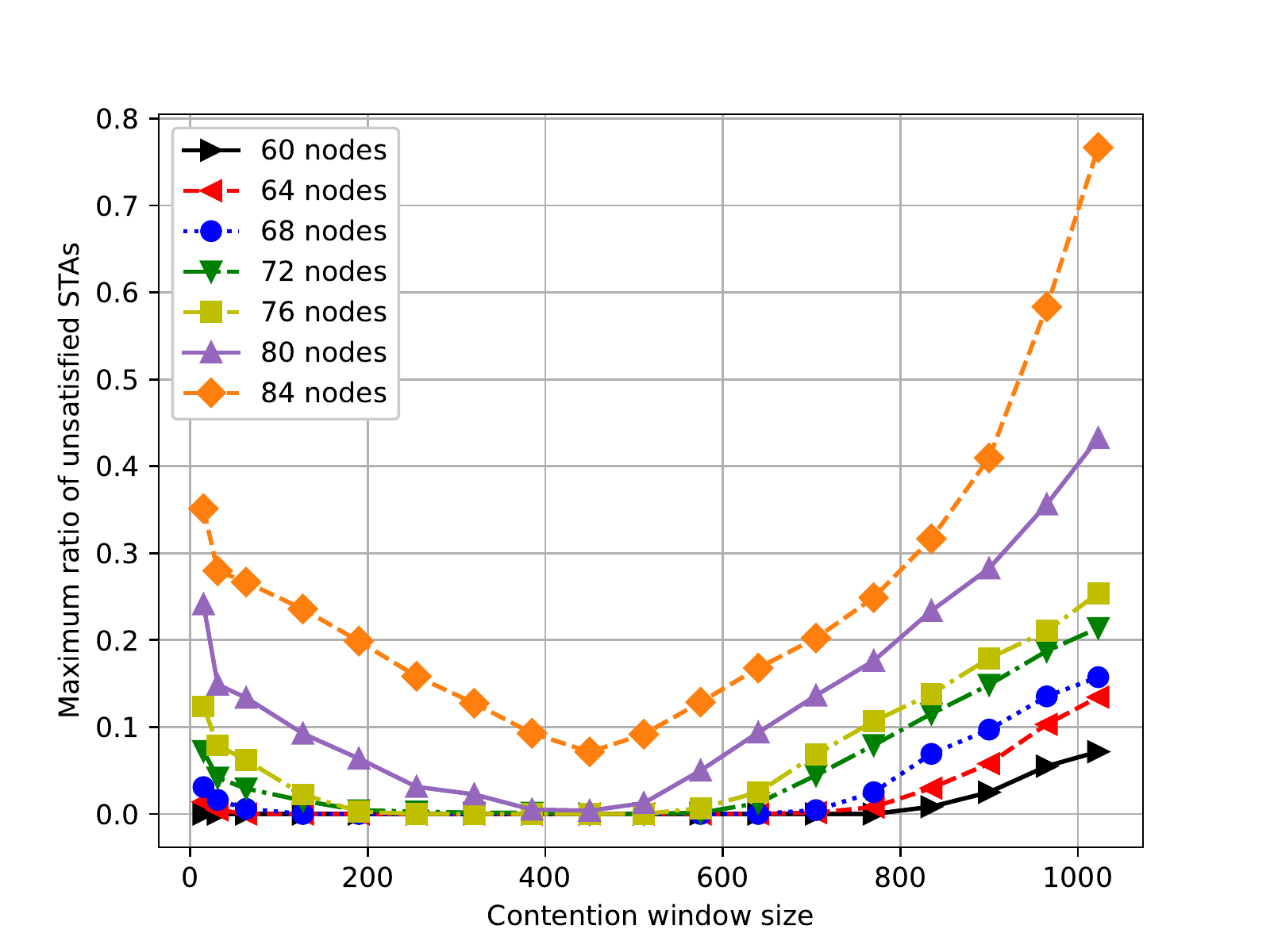} \\ (b)}
	\end{minipage}
	\caption{\label{fig:best_cw} The maximum ratio of unsatisfied STAs for the IEEE~802.11n channel bonding technique (a) in the symmetric case (b) in the asymmetric case.}
\end{figure}

To sum up, the presented results reveal that the channel bonding technique from IEEE~802.11n outperforms other channel access methods. Furthermore, as IEEE~802.11bd permits using only $CW\in \{15, 31, 63, 127, 255, 511, 1023\}$, we recommend using $CW = 511$ for the legacy technique because this value provides the unsatisfied STAs ratio below $10\%$ with the higher number of vehicles compared with other $CW$ values from the set, see Fig.~\ref{fig:best_cw}. In particular, $CW = 511$ can provide 4 times decrease in the unsatisfied STAs ratio compared with default $CW = 15$.

\section{Conclusion}
\label{sec:conclusion}
In this study, we performed an extensive comparative analysis of channel access methods, which can be used in IEEE~802.11bd networks: single-channel access method from IEEE~802.11p, the legacy channel bonding technique from IEEE~802.11n, and two techniques described in the IEEE~802.11bd draft. Moreover, we considered different contention window sizes.

Although in many scenarios the performance of various channel bonding techniques is close to each other, the IEEE 802.11n technique ensures better quality of service in terms of frame transmission delays and packet loss ratio compared with other channel access methods. Furthermore, we show that in case of losses, the performance is very sensitive to the contention window and we propose to increase the contention size to 511 because it can provide a four times lower unsatisfied STAs ratio comparing with the default contention window size on heavy loaded roads. 

\bibliographystyle{IEEEtran}
\bibliography{refs-new}

\end{document}